\documentclass[aps,pra,reprint,groupedaddress,longbibliography]{revtex4-1}
\usepackage{physics}
\usepackage{multirow}
\usepackage{graphicx}
\usepackage{bbm}
\usepackage{bbold}

\begin{document}


\title{High-fidelity measurement of qubits encoded in multilevel superconducting circuits}


\newcommand{\eqcontrib}{These authors contributed equally to this work. Email: sal.elder@yale.edu; christopher.wang@yale.edu}
\author{Salvatore S. Elder}
\thanks{\eqcontrib}
\author{Christopher S. Wang}
\thanks{\eqcontrib}
\author{Philip Reinhold}
\author{Connor T. Hann}\author{Kevin~S.~Chou}
\altaffiliation{Current address: Quantum Circuits, Inc., New Haven, CT 06520, USA}
\author{Brian~J.~Lester}
\altaffiliation{Current address: Atom Computing, Berkeley, CA 94710, USA}
\author{Serge Rosenblum}
\altaffiliation{Current address: Department of Condensed Matter Physics, Weizmann Institute of Science, Rehovot, Israel}\author{Luigi Frunzio}\author{Liang Jiang}
\altaffiliation{Current address: Pritzker School of Molecular Engineering, University of Chicago, Chicago, IL 60637, USA}\author{Robert J. Schoelkopf}\email{Email: robert.schoelkopf@yale.edu}
\affiliation{Department of Applied Physics and Physics, Yale University, New Haven, CT 06520, USA}
\affiliation{Yale Quantum Institute, Yale University, New Haven, CT, USA}


\date{\today}

\begin{abstract}
Qubit measurements are central to quantum information processing.
In the field of superconducting qubits, standard readout techniques are not only limited by the signal-to-noise ratio, but also by state relaxation during the measurement.
In this work, we demonstrate that the limitation due to relaxation can be suppressed by using the many-level Hilbert space of superconducting circuits: in a multilevel encoding, the measurement is only corrupted when multiple errors occur.
Employing this technique, we show that we can directly resolve transmon gate errors at the level of one part in $10^3.$
Extending this idea, we apply the same principles to the measurement of a logical qubit encoded in a bosonic mode and detected with a transmon ancilla, implementing a proposal by Hann et al. [Phys. Rev. A \textbf{98} 022305 (2018)].
Qubit state assignments are made based on a sequence of repeated readouts, further reducing the overall infidelity. This approach is quite general and several encodings are studied; the codewords are more distinguishable when the distance between them is increased with respect to photon loss.
The tradeoff between multiple readouts and state relaxation is explored and shown to be consistent with the photon-loss model. 
We report a logical assignment infidelity of $5.8\times 10^{-5}$ for a Fock-based encoding and $4.2\times 10^{-3}$ for a QEC code (the $S=2,N=1$ binomial code). Our results will not only improve the fidelity of quantum information applications, but also enable more precise characterization of process or gate errors.
\end{abstract}


\maketitle


\section{Introduction}
Quantum information processing (QIP) involves many tasks. One requirement, crucial to any QIP experiment, is the ability to \emph{measure} a qubit or qubit register. Given a superposition $\ket{\psi}=\alpha\ket{0_L}+\beta\ket{1_L},$ our task is to collapse the state and detect the corresponding classical bit of information. That is, we should extract a ``0'' with probability $|\alpha|^2,$ and a ``1'' with probability $|\beta|^2,$ projecting onto the state $\ket{0_L}$ or $\ket{1_L},$ respectively. A canonical example of qubit measurement is at the end of a quantum computation, after which the experimenter measures the qubit array and infers a useful result. Experimental implementations of quantum computing requirements, including feedback-based state preparation \cite{article:twoiongate}, gate calibration, and error-syndrome extraction for quantum error correction \cite{article:shor, article:kl, article:ion_qec, article:cat_experiment, article:ft} (QEC), also rely on qubit measurements. Outside of quantum computing applications, measurement is used in quantum key distribution \cite{proc:qkd}, enhanced sensing \cite{article:metrology_review}, teleportation of states or gates \cite{article:teleportation_theory, article:teleportation_wineland, article:chou}, and fundamental tests of locality and entanglement \cite{article:aspect1, article:aspect2}. Although the particular metrics will depend on the application at hand, better measurements will translate into better results in all of the examples mentioned. Furthermore, improvements in qubit measurement will improve the calibration and characterization \cite{article:rb} of gates and other quantum operations. Finally, we point out that in order to implement large quantum circuits, measurements must improve alongside advances in gate quality and number of qubits. For example, for fixed single-qubit measurement error, the probability that an entire register of qubits will be measured correctly decreases exponentially in the size of the register.

Much progress has been made in the experimental implementation of single-shot qubit measurements in a variety of systems. Single-shot measurements of qubits based on trapped ions \cite{article:ion_hume, article:ion_myerson, article:ion_seif}, electron spins in quantum dots \cite{article:enhanced_spin_readout, article:si_spin_readout, article:spin_qubit_review}, and superconducting circuits \cite{article:wallraff, article:walter} have all been demonstrated with fidelities higher than 99\%.
Improvements have been made in a variety of ways; interesting approaches have included mapping a spin state onto a metastable charge state \cite{article:enhanced_spin_readout}, and the repeated readout of an ion state using an ancilla \cite{article:ion_hume}.

In all of the systems mentioned above, state relaxation is a major limitation to the measurement fidelity. In a continuous measurement, some signal (such as photomultiplier current or the phase of a microwave tone) is recorded and compared to the expected response corresponding to each qubit state. Because noise causes uncertainty in the assignment of individual outcomes, we would generally like to acquire signal for a long time in order to improve the measurement contrast.
On the other hand, $T_1$ events during the measurement, in which the qubit relaxes from its excited to ground state, lead to incorrect assignment of the initial state \cite{article:gambetta}.

In this work, we overcome the first-order $T_1$ limit by encoding qubits in multilevel systems. If the physical states representing  the $0$ and $1$ bit are separated by multiple energy levels, then a single relaxation event will \emph{not} corrupt the 0 or 1 which was encoded. For this reason, qubit encodings with a larger distance between codewords with respect to the dominant error channel can be measured with much improved fidelity.

As a simple and direct example of our approach, we study the measurement of a transmon qubit $(\hat{t})$ dispersively coupled \cite{article:dispersive} to a readout resonator $(\hat{r})$.
In particular, using higher levels of the transmon is shown to improve the fidelity by two orders of magnitude.
In Sec.~\ref{sec:oscillator}, in order to more fully demonstrate the advantage of multilevel encodings, we encode a qubit in a harmonic oscillator implemented as a $\lambda/4$ coaxial superconducting cavity. Using a bosonic mode allows us to systematically explore different qubit encodings, including Fock-based encodings as well as error-correctable binomial codes \cite{article:binomial}. The information in this ``storage'' mode ($\hat{s}$) is read out using the dispersively coupled transmon as an ancilla according to a recent proposal \cite{article:hann}. In this proposal, the storage-ancilla interaction is used to map information from the storage onto the ancilla, and the ancilla-readout interaction is used to read out the ancilla state.

\section{Background}
\begin{figure}
\includegraphics{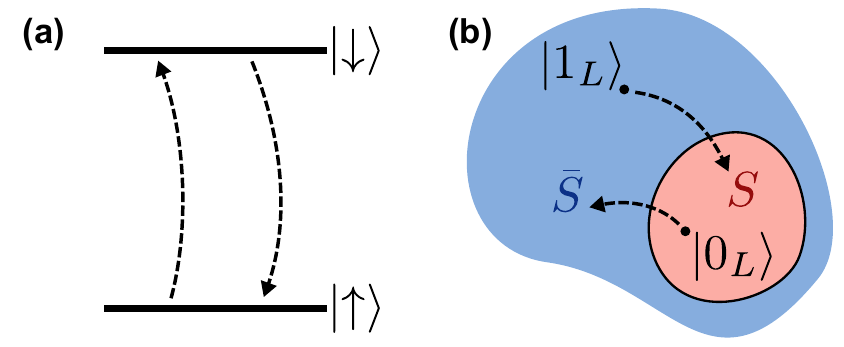}
\caption{\textbf{Cartoon illustrating qubit readout errors.} \textbf{(a)} In a two-level system, qubit readout means determining the energy eigenstate of the system at time $t=0.$ Dashed arrows indicate incorrect assignments, $P(\uparrow | \ket{\downarrow})$ and $P(\downarrow|\ket{\uparrow}),$ which may be due to detector noise or state transitions. \textbf{(b)} In a many-dimensional Hilbert space, reading out a single bit requires partitioning Hilbert space into two sets containing the encoded 0 and 1 states. Instead of an energy eigenstate, membership in $S$ or its complement $\bar{S}=\mathcal{H}\setminus S$ is measured. The dashed arrows now indicate the assignment errors $P(S|\ket{1_L})$ and $P(\bar{S}|\ket{0_L}).$}
\label{fig:subspaces}
\end{figure}

Before going further, it is worth clarifying what is meant by encoding and measuring a bit in a multilevel system. Traditionally, a qubit is identified with a two-level system such as a spin-$\frac{1}{2}$ particle in a magnetic field. In this case, there is no trouble identifying the two eigenstates with the values of a bit: $\ket{\uparrow}$ represents a 0, say, and $\ket{\downarrow}$ represents a 1. To measure the bit encoded by such a system means performing a projective measurement of an eigenstate of the system. Finally, to incorrectly measure the bit encoded by such a system means that state transitions or detector noise cause the experimenter to record a state initially $\ket{\downarrow}$ as ``0'' or vice-versa. These simple ideas are sketched in Fig.~\ref{fig:subspaces}(a), where dashed arrows indicate incorrect measurements.

Actual implementations of qubits, however, may contain more than two energy levels. In this case, the definition of a qubit is a matter of convention. For example, the transmon \cite{article:transmon,article:transmon_exp} is an anharmonic oscillator with several energy levels ($\ket{g},$ $\ket{e},$ $\ket{f},$ $\ket{h},$ \ldots) in which the ground and first excited state are often operated as a qubit. This is possible because the $\ket{g}$--$\ket{e}$ transition is detuned by many linewidths from all other transitions, so that it can be individually addressed by a single rf drive. In this paper we consider representing and measuring information encoded in the higher states.

Even more care is required when dealing with a harmonic oscillator, in which many transitions fall within the mode linewidth and there is no obvious notion of a qubit. In fact, several continuous-variable codes \cite{article:gkp, article:cat_theory, article:binomial, article:victor, misc:linshu, article:chuang18, article:chuang97, article:cat99} have been studied, in which two states in Hilbert space, $\ket{0_L}$ and $\ket{1_L},$ are designated as the logically encoded 0 and 1 states. What does it mean to \emph{measure} qubits encoded in this way? To answer this question, 
we imagine partitioning the Hilbert space $\mathcal{H}$ into two subsets, $S$ and $\bar{S},$ containing $\ket{0_L}$ and $\ket{1_L}$ respectively. An ideal measurement of the encoded bit is a projective measurement of the \emph{membership in $S.$} The POVM operators \cite{book:mikeike} corresponding to such a measurement are

\begin{align}
\label{eq:povm}
\mathcal{M}_0&=\sum_{\ket{\psi}\in S}\ket{\psi}\bra{\psi},\\
\mathcal{M}_1&= \mathbb{1}-\mathcal{M}_0.
\end{align}
Finally, the \emph{fidelity} of such a measurement is defined \cite{article:gambetta} as
\begin{equation}
F=1-P\Big(S\Big | \ket{1_L}\Big)-P\Big(\bar{S}\Big| \ket{0_L} \Big),
\end{equation}
where $P(a|s)$ is the probability that an assignment $a$ will be made, given that the true initial state was $s.$ Note that this definition goes to zero, not $0.5,$ for random guessing. The fidelity offers a simple, unified metric with which we can compare the measurement of different codes.

\section{Multilevel Transmon Measurement}
\label{sec:transmon}
In the standard approach to superconducting qubit measurement, a readout resonator is dispersively coupled to a transmon, meaning that the frequency of the resonator mode is shifted if the transmon is in an excited state. By driving the resonator and recording the amplitude and phase of the response, one can therefore infer the transmon state. Of course, there is inevitably noise on top of the average resonator response. Integrating the signal for a longer time provides more separation between the two averages.
However, transmons undergo random transitions between eigenstates with a characteristic time $T_1.$ After a certain fraction of $T_1,$ the measurement signal will no longer provide useful information about the initial state; therefore, the signal-to-noise ratio competes with the lifetime of the information being measured \cite{article:gambetta}.

Suppose that instead of asking whether the qubit was initially in its ground state $\ket{g}$ or first excited state $\ket{e},$ we ask a slightly different question: Was the qubit initially in its ground state $\ket{g},$ or in an excited state $\ket{e}, \ket{f}, \ket{h},\ldots$? Extending the logic above, the $n$th excited state will remain distinguishable from $\ket{g}$ until $n$ relaxation events have occured. Therefore we are able to acquire signal for a longer time, leading to a better signal-to-noise ratio. Additionally, the figure of merit $t_\mathrm{m}/T_1\ll 1$ for relaxation to the ground state becomes $(t_\mathrm{m}/T_1)^n$ when discriminating between $\ket{g}$ and the $n$th excited state, where $t_\mathrm{m}$ is the measurement acquisition time
\footnote{An anharmonic oscillator is not necessarily characterized by a single $T_1$ value \cite{inbook:spectrometers}. In this context the expression $(t_\mathrm{m}/T_1)^n$ must be regarded as a rule of thumb and not an exact value.}. This enables us to perform high-fidelity measurements of transmon states.
The measurement error for a related system was previously studied as a function of the signal-to-noise ratio and the number of excited states \cite{article:subpoisson}.

\begin{figure*}
\center
\includegraphics{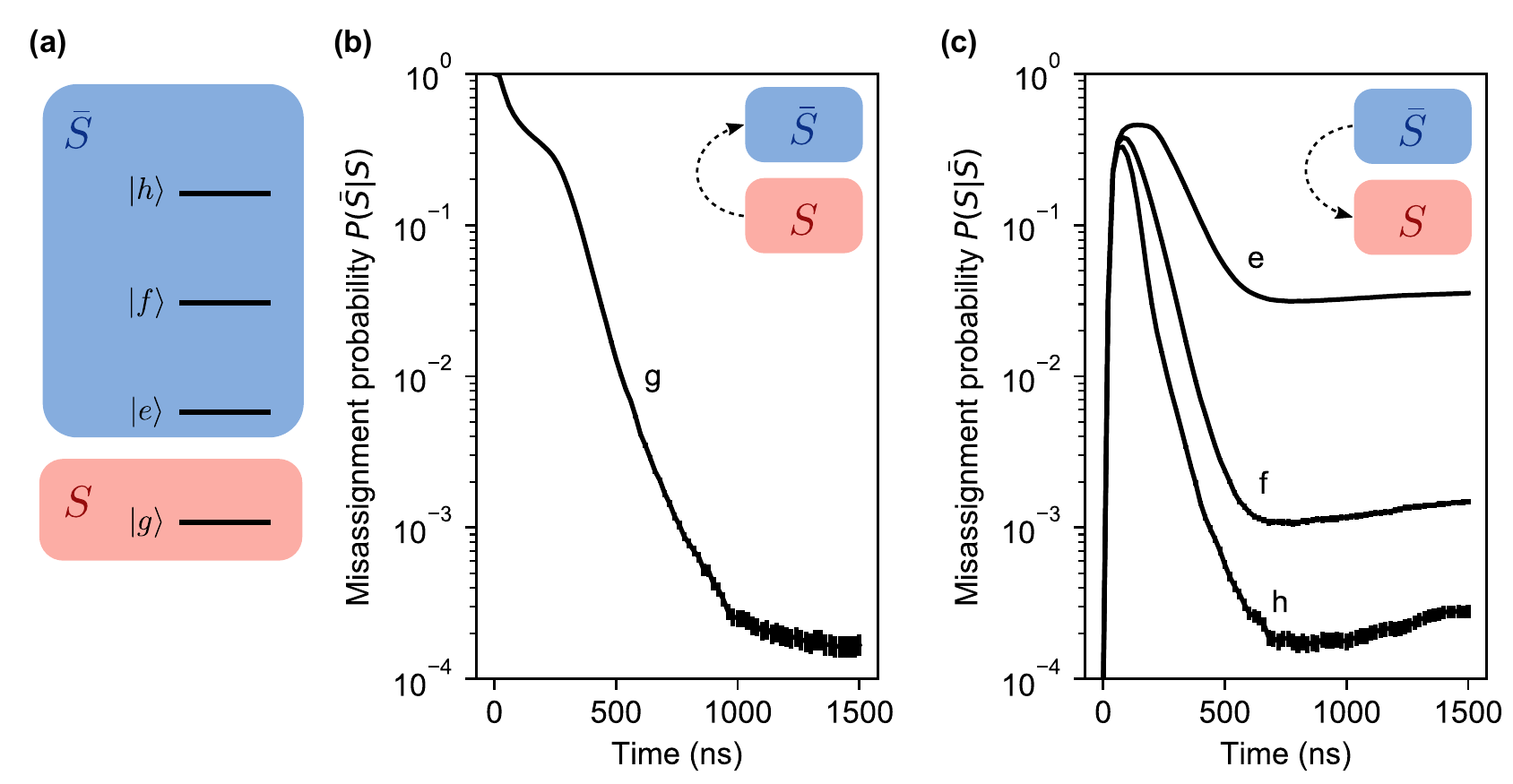}
\caption{\label{fig:transmon_buffer} \textbf{High-fidelity measurement of a transmon.} \textbf{(a)} Representation of the measurement performed. The first four states of the transmon are individually prepared, measured, and classified as belonging to $S$ or $\bar{S}$ based on the measurement record. \textbf{(b)} The probability that the ground state $\ket{g}$ is incorrectly assigned as belonging to $\bar{S},$ plotted as a function of the length of measurement signal used to make the classification. The initial shape of the curve is related to the ring-up time of the readout resonator. The misassignment probability decreases as a function of time, because collecting more signal improves the separation of readout signals. However, the improvement stops once the misassignment probability is comparable to the probability that the transmon has gained a photon. Collecting more signal would cause additional misassignments. \textbf{(c)} The probability that an excited state $\ket{e},$ $\ket{f},$ or $\ket{h}$ is incorrectly assigned as beloning to $S.$ Again, after some initial transient behaviour, the misassignment probability improves as signal is collected for a longer time. Eventually, relaxation events cause misassignments, and the curves increase for large acquisition times. The probability that all excitations are lost, causing erroneous assignment to $S,$ is much smaller when a higher excited state is prepared. The exact misassignment probabilities achieved depend on many factors, especially the relaxation time $T_1^{(ge)}\approx 51~\mu$s and thermal population $\bar{n}_t\approx 0.4\%.$}
\end{figure*}

In order to test or characterize a high-fidelity transmon measurement, we first need to prepare the states $\ket{g},\ \ket{e},\ \ket{f},$ and $\ket{h}$ accurately. This is done by applying the appropriate pulse sequence, then performing a preliminary ``check'' measurement and postselecting on measurements which give us a high confidence that the correct state has been prepared. Because we record all outcomes following a successful state preparation, this method is a fair way to characterize the measurement fidelity \footnote{This method of state preparation is analogous to the one described in the supplemental material.}.

We test the fidelity by measuring the transmon and comparing the resulting assignment to the state which was prepared.
The resulting misassignment probabilities are plotted in Fig.~\ref{fig:transmon_buffer} as functions of the measurement time.
To classify an individual measurement record $z(t)$ up to a particular time $t_\mathrm{m},$ the record is compared to the average signals, $\bar{z}_g(t),\ldots, \bar{z}_h(t)$ corresponding to the $\ket{g},\ldots, \ket{h}$ states. The classifier outputs the state $s$ for which $\sum_{t=0}^{t_\mathrm{m}}\left|z(t)-z_s(t)\right|^2$ is minimal. These labels are used to determine whether a particular realization was assigned correctly as either the ground state or an excited state.

We can understand the shapes of these plots qualitatively as follows: in the first part of the measurement, we acquire more signal and the misassignment probabilities decrease. Eventually, however, we are limited by state relaxation and the measurement no longer improves. If we continue to collect signal and classify the states naively, then we will actually start to make more misassignments due to the state relaxation. The probability that state $\ket{e},$ $\ket{f},$ or $\ket{h}$ decays to $\ket{g}$ has $t^1, t^2,$ or $t^3$ dependence, leading to the improvement in measurement fidelity shown in Fig.~\ref{fig:transmon_buffer}(c). We observe that if a transmon is prepared in the $\ket{g}$--$\ket{h}$ manifold, then it can be read out with a measurement infidelity of $P(S\mid \ket{h})+P(\bar{S}\mid \ket{g})=(4.0 \pm 0.5)\times 10^{-4}.$

\begin{figure}
\includegraphics{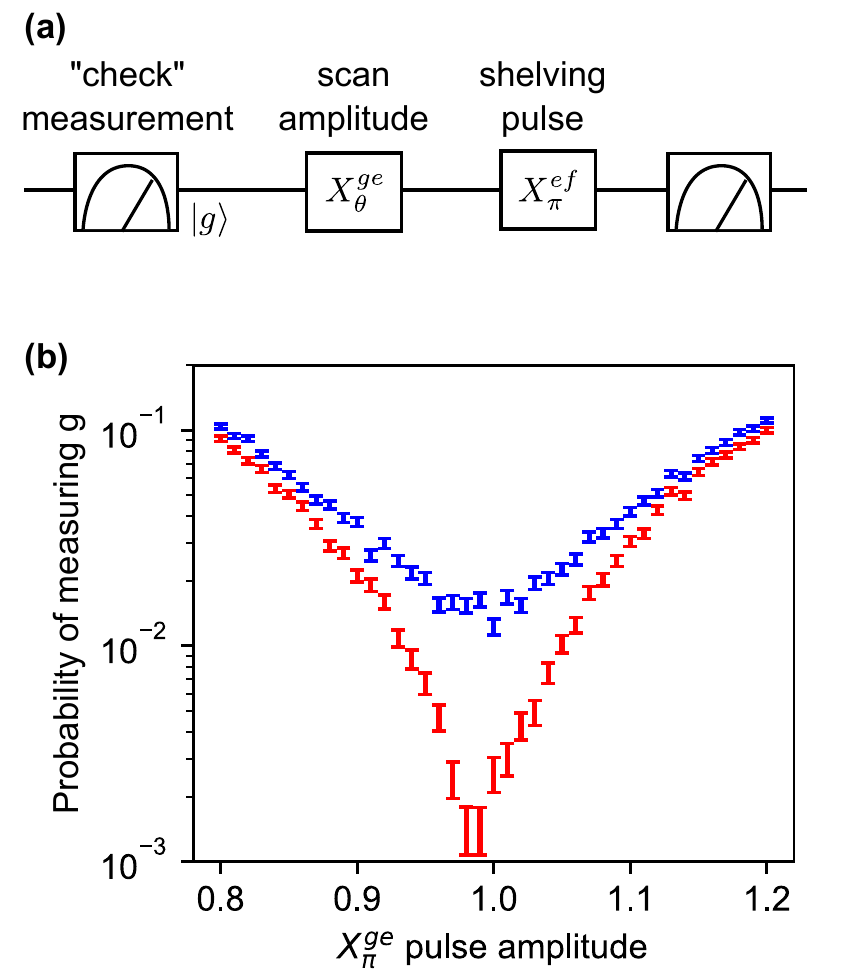}
\caption{\label{fig:rabi_fidelity} \textbf{Improved calibration and characterization of gate error.} \textbf{(a)} Schematic and \textbf{(b)} results of the shelved measurement demonstration. After the ground state is prepared, a $g$--$e$ $\pi$-pulse is applied with variable amplitude and the transmon is measured. An $e$--$f$ ``shelving'' pulse before the measurement greatly improves the visibility of $g$--$e$ Rabi oscillation. The unshelved measurement is shown in blue, and the shelved measurement is shown in red.}
\end{figure}

We can apply these ideas to the more usual task of measuring ``g'' or ``e'' using a shelving technique. The idea is simple: we apply a rapid $e$--$f$ $\pi$-pulse, $X_\pi^{ef}=\ket{f}\bra{e}+\ket{e}\bra{f},$ to the transmon immediately before measuring it. In this way, we transform the problem of distinguishing $\ket{g}$ from $\ket{e}$ into the problem of distinguishing $\ket{g}$ from $\ket{f},$ which
can be done with higher fidelity. This has been previously explored as a method of increasing the contrast of qubit measurements with a latching readout \cite{article:cqed_shelving}. We observe that in such a scheme, the misassignment of $\ket{e}$ as ``g'' is a second-order error. On the other hand, the misassignment probability is directly sensitive to preparation error. Therefore, if $\ket{e}$ is prepared with a pulse, then the error in that pulse will limit the performance of the shelved measurement.

However, this means that a shelved measurement has a resolution comparable to the transmon gate errors. A measurement with such resolution is highly desirable in circuit QED, in which single-qubit gate errors are typically obscured by much larger measurement errors.
We now show, using the above method for improved $\ket{g}$--$\ket{e}$ readout, how one can calibrate gates in a way that reduces state preparation and measurement (SPAM) errors. For example, a $g$--$e$ $\pi$-pulse, $X_\pi^{ge}=\ket{g}\bra{e}+\ket{e}\bra{g},$ would typically be calibrated by the following procedure. (1) Prepare the transmon in its ground state, $\ket{g}.$ (2) Apply the pulse with a variable amplitude. (3) Measure the state of the qubit. The result of such a procedure is shown in Fig.~\ref{fig:rabi_fidelity} in blue. The minimum value of this curve, $(1.2\pm 0.1)\times 10^{-2},$ is the inferred gate error, uncorrected for SPAM errors. Alternatively, the same procedure can be performed using shelving as described above. As shown in red in Fig.~\ref{fig:rabi_fidelity}, the visibility is greatly improved because of the reduced measurement error. We therefore obtain, by direct measurement, an improved bound of $(1.4\pm 0.3)\times 10^{-3}$ on the gate error.

We have shown how to characterize a gate to within a small factor.
A QuTiP \cite{article:qutip} simulation \footnote{See the supplemental material.} predicts a residual $g$ population of $P_g=3.7\times 10^{-4}$ after the $g$--$e$ $\pi$-pulse. This is the quantity we would like to be able to measure directly. Similar calculations predict $P_g^\textrm{shelved}=7.2\times 10^{-4}$ at the end of the $e$--$f$ pulse, which is somewhat larger due to additional relaxation during the $e$--$f$ pulse. Finally, in order to roughly estimate the probability of reading out ``g,'' we calculate $P_g^\textrm{meas}=9.7\times 10^{-4}$ halfway through the measurement. These calculations show that although there is some error in the shelved measurement itself, it is comparable to the pulse error we wish to characterize.
We stress that if the gate fidelity ($\sim t_\mathrm{gate}/T_2$) were to improve, then the shelved measurement would improve with it.

\section{Measurement of Logically Encoded Qubits}
\label{sec:oscillator}
As mentioned earlier, continuous-variable (CV) encodings of quantum information in bosonic modes offer a promising route to fault-tolerant quantum computation \cite{article:gkp, article:cat_njp, article:victor}. Instead of encoding a logical qubit in many physical qubits, CV schemes use an infinite-dimensional Hilbert space to build the redundancy needed for error correction into a single oscillator or storage mode. In this section, we show how redundancy and distance can be used to improve logical qubit measurements.
We implement a proposal \cite{article:hann} to measure multiphoton encodings in a harmonic storage mode with high fidelity. The method consists of mapping the encoded information onto an ancilla, reading out the ancilla, resetting the ancilla, and repeating these steps several times.

To understand the advantage of a repeated readout scheme, first consider the error associated with mapping information onto the ancilla and reading it out. If this process is noisy, it is liable to give the wrong answer, but if it is QND with respect to the state of the storage mode---that is, if the readout does not induce extra transitions between eigenstates---then the storage mode can be read out repeatedly. To make an assignment of the bit encoded by the storage mode, one could take a majority vote of the outcomes obtained via the ancilla. In this way, many individual \emph{readouts} of the encoded information can be combined to a single \emph{measurement} with a greatly reduced probability of error. Next consider the relaxation of the storage mode itself. As photons are lost incoherently from the storage, the information encoded by the storage is corrupted, limiting the ability of any measurement to extract the initially encoded bit. For example, if information were encoded in Fock states $\ket{0}$ and $\ket{1},$ then the relaxation rate, $\kappa_\downarrow,$ times the total mapping, readout, and reset time, $\tau,$ would (with the appropriate prefactor) set a lower bound on the probability of measurement error. On the other hand, if the logical codewords were separated by $L$ Fock states, then the measurement would be robust up to $L$ relaxation events. Therefore the bound would scale like $(\kappa_\downarrow \tau)^L,$ an exponential improvement.

The measurement infidelity has been estimated for various multiphoton encodings which are repeatedly read out using an ancilla as described above. It was shown \cite{article:hann} that the measurement infidelity can be broken into two contributions, corresponding to the observations above: first, the probability that a majority of individual readouts will give the wrong answer (ignoring state relaxation), and second, the probability that the state will relax from $S$ into $\bar{S}$ (or vice versa) during the first half of the repeated readout sequence. These terms are given by
\begin{align}
\label{eq:theory}
1-F=&L\left(\left\lceil \frac{N}{2}\right\rceil \kappa_\downarrow\tau \right)^{L-1}+\left(\left\lceil\frac{N}{2}\right\rceil \kappa_\uparrow \tau\right)^2 \\
+&\binom{N}{\lceil N/2\rceil}\delta_0^{\lceil N/2\rceil} + \binom{N}{\lceil N/2\rceil}\delta_1^{\lceil N/2\rceil}\nonumber
\end{align}
for the family of Fock codes $\ket{0_L}=\ket{0},$ $\ket{1_L}=\ket{L}.$ In the equation above, $N$ denotes the number of measurements made, $\kappa_{\downarrow(\uparrow)}$ is the rate of energy loss (gain) in the storage mode, and $\delta_{0(1)}$ is the probability of a readout error during a single round of measurement when the $\ket{0_L}$ ($\ket{1_L}$) state was prepared. We choose $S=\mathrm{span}\{\ket{0},\ket{1}\}$ to measure this encoding, so that the measurement of the $\ket{0}$ state is robust to the gain of a single photon.

\begin{figure*}
\includegraphics{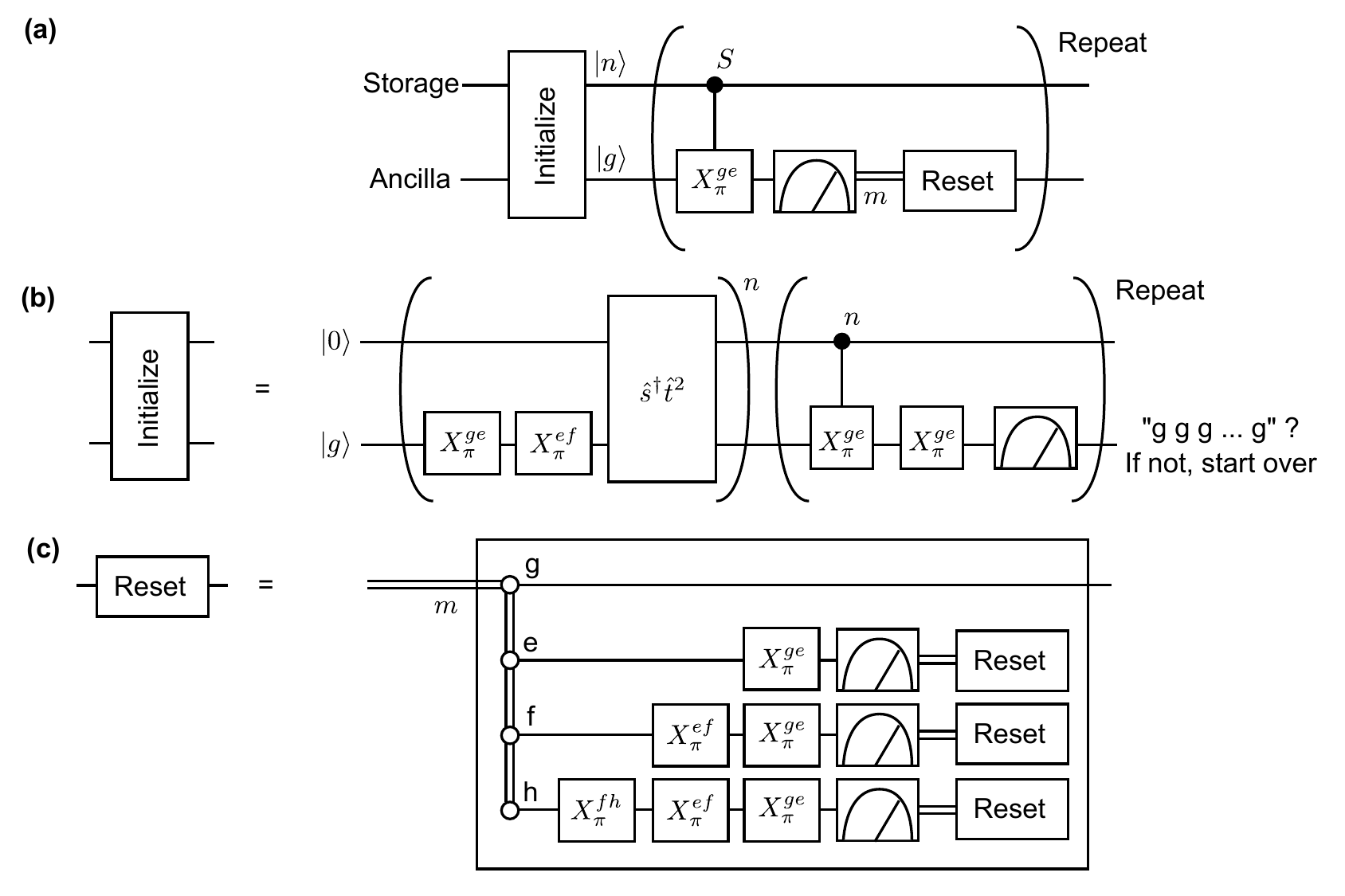}
\caption{\label{fig:schematics} \textbf{Schematic of repeated measurement procedure.} \textbf{(a)} Block diagram showing how an encoded bit is measured. We first prepare one of several Fock states $\ket{n}$ in the storage mode. We then perform several readout cycles, each time obtaining an outcome. Each cycle consists of a decode pulse which excites the ancilla conditioned on the encoded bit, followed by a readout and reset of the ancilla. \textbf{(b)} The initialization procedure uses a series of number-selective pulses to verify that the correct state has been prepared. This verification is crucial in order to demonstrate a sensitive detection. \textbf{(c)} Each reset consists of a real-time feedforward protocol which ensures that the ancilla is in its ground state $\ket{g}$ at the end of each cycle.}

\end{figure*}

The measurement protocol described allows us to perform many independent readouts of an encoded qubit using only a single ancilla. This is accomplished using an adaptive feedforward scheme, which is illustrated in Fig.~\ref{fig:schematics}. A high-level description is given in 4(a), with further implementation details shown in 4(b) and 4(c). We next discuss the experimental implementation of the protocol.

Once the system is initialized, the first step of the measurement is to map the information encoded in the storage mode onto the state of the ancilla. This step is shown as an $S$--controlled $X_\pi^{ge}$ gate in Fig.~\ref{fig:schematics}(a). The realization of this entangling gate is based on the dispersive interaction between $\hat{s}$ and $\hat{t},$ which imparts an ancilla \emph{frequency} shift which depends on the \emph{number} of excitations in the storage mode:
\begin{align}\hat{H}/\hbar&=\omega_t \hat{t}^\dagger \hat{t}+\chi_{st}\hat{s}^\dagger \hat{s}^\dagger \hat{t}^\dagger \hat{t}+\cdots\\
&=\left(\omega_t+\chi_{st}\hat{s}^\dagger \hat{s}\right)\hat{t}^\dagger \hat{t}+\cdots\nonumber\end{align}
An ancilla pulse with spectral content near the frequencies corresponding to the $n$ in $S$ has the effect of flipping the ancilla state if and only if the storage state is in $S.$ For example, when measuring the Fock codes, the ancilla is excited if the storage is in the state $\ket{0}$ or $\ket{1}.$

After the information encoded by the storage mode has been mapped onto the ancilla, the state of the ancilla is read out. The outcome $m$ provides information about the encoded qubit state, and provides the repeated measurement protocol with one vote.

However, the outcome of the readout also determines the operation which must be performed in order to reset the ancilla. The readout was calibrated to distinguish between the states $\ket{g},$ $\ket{e},$ $\ket{f},$ and $\ket{h}.$ Ideally the ancilla would stay in the $\ket{g}$--$\ket{e}$ manifold, but resolving higher states enables a more robust reset operation. The reset protocol is shown in Fig.~\ref{fig:schematics}(c) as a recursive block diagram. It relies on realtime logic implemented on FPGA cards, and does not terminate until it successfully records the ancilla in its ground state $\ket{g}.$ Once the ancilla is reset to its ground state, it is available for additional measurements. In this way, the map--measure--reset cycle is repeated many times in order to extract a single high-fidelity measurement of the logical qubit.

In order to demonstrate a high-fidelity measurement of the storage mode, it is crucial that the initialization step prepare states accurately. As shown in Fig.~\ref{fig:schematics}(b), number states were prepared by creating two excitations in the ancilla, then using a sideband interaction to convert them into a storage excitation \cite{article:0f1g}, and repeating this process the desired number of times. This process is associated with a significant initialization error (on the order of $\sim 10\%$ overall when preparing states of several photons), which would dominate over the measurement infidelity and prevent us from making any conclusions about the performance of our protocol. Therefore, after the state creation routine is finished, number-selective pulses and ancilla readouts are used to repeatedly check that the correct state was prepared. Only if several checks pass is the state preparation accepted.

\begin{figure*}
\includegraphics{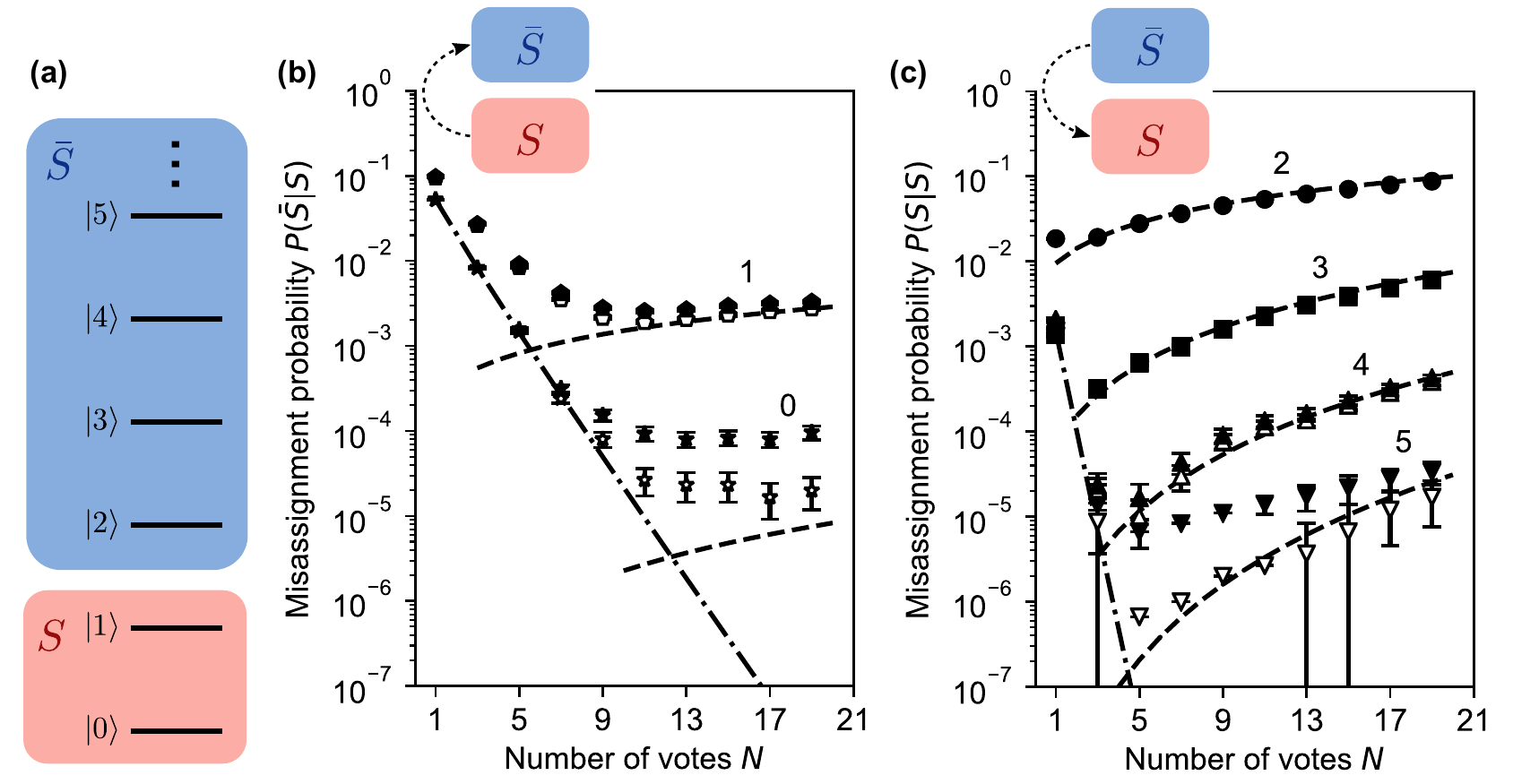}
\caption{\label{fig:votes} \textbf{Enhancement of measurement fidelity with code distance.} \textbf{(a)} In these experiments, we excite the ancilla if the storage mode has fewer than two photons in it. In doing so, we read out whether the storage state is in $S$ or $\bar{S}$ as shown. \textbf{(b)} Probability that when a state in $S$ is prepared in the storage mode, it is assigned incorrectly as $\bar{S}.$ The state $\ket{0}$ or $\ket{1}$ is prepared in the storage mode, as indicated by the labels on the plot. 
The data show that the fidelity improves exponentially with the number of votes, until excitation out of the $S$ subspace limits the measurement. Dash-dotted lines show the terms in the theoretical prediction corresponding to majority voting, dashed lines show terms corresponding to state transitions, and closed symbols show the experimental data. Open symbols indicate postselection on successful ancilla resets.
\textbf{(c)} Probability that when a state in $\bar{S}$ is prepared in the storage mode, it is assigned incorrectly as $S.$ Again, we see that majority voting suppresses the misassignment probability, but that for large $N$ the misassignment probability increases due to photon loss. States of higher photon number are measured with smaller error because the measurement is robust to more photon loss events.}
\end{figure*}

The results of a repeated measurement experiment are plotted in Fig.~\ref{fig:votes} and illustrate all of the expected behavior. Fig.~\ref{fig:votes}(a) defines the operator which is measured; in this case the goal is to measure whether the storage was initially in the state $\ket{0}$ or $\ket{L}$ for some $L\geq 2.$ We plot the results of a majority vote as a function of the number of readouts taken. Furthermore, the assignment errors are split into two types: $\ket{0_L}$ incorrectly assigned as $1,$ and vice-versa. As more readouts are incorporated into the majority vote, the measurement fidelity improves because mapping and readout errors are suppressed. Eventually, however, this suppression competes with state transitions in the storage mode itself, causing the measurement to degrade once too many readouts are taken. We also observe that as the distance between codewords ($L$) increases, the measurement fidelity improves dramatically.
These trends are captured by their theoretical description \cite{article:hann} and follow the predictions of Eq.~\eqref{eq:theory}, shown in dashed and dash-dotted lines.

\begin{figure}[h]
\includegraphics{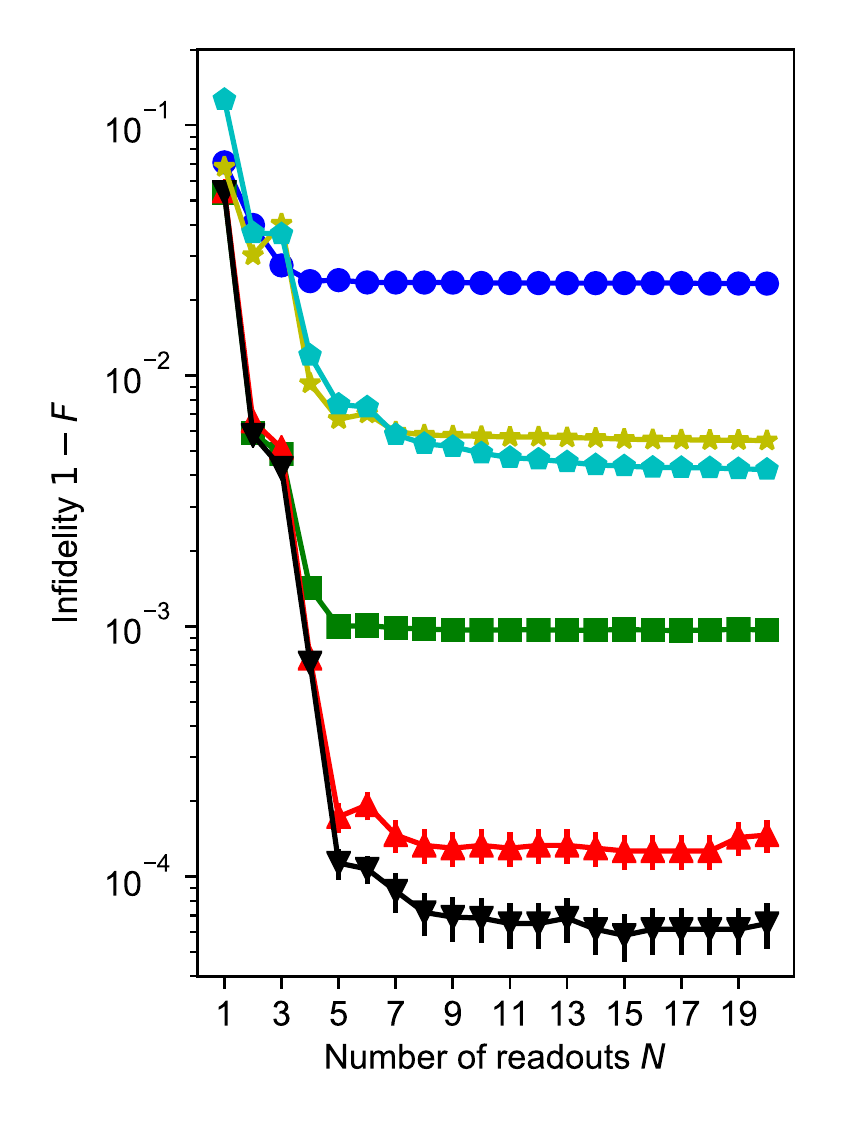}
\caption{\label{fig:mle} \textbf{High-fidelity measurement of bosonic codes.} Results of measuring the logical $Z$ observable in different qubit encodings. Given single-shot ancilla assignments, a Bayesian classifier outputs either ``$0_L$'' or ``$1_L$,'' according to its best estimate of the initial state. As more readouts are incorporated into the estimate, the measurement infidelity decreases monotonically. Results for the 0--2, 0--3, 0--4, and 0--5 Fock codes are shown in blue circles, green squares, red triangles, and black triangles, respectively. Results for the first- and second-order binomial codes are shown in yellow stars and cyan pentagons. As the code distance increases, the measurement fidelity improves.}
\end{figure}

The open symbols in Fig.~\ref{fig:votes} represent postselection of successful reset operations. By removing 0.2\% of records corresponding to nonideal ancilla resets, we obtain much better agreement with the theoretical predictions. These events appear to be due to rare excitations to high levels in the ancilla, the origin of which is unclear \footnote{See the supplemental material for a detailed discussion of ancilla reset errors.}. It is worth pointing out that we do this postselection only in order to compare to the theory, not to make fair qubit measurements. We emphasize that the results in the remainder of this paper, in particular Fig.~\ref{fig:mle} and Table~\ref{tab:codes}, are \emph{not} postselected in this way. They represent ``fair'' measurements.

\begin{table*}
\begin{tabular}{cc@{\hspace{1cm}}cccc}
\hline\hline
Code      & Flip ancilla if $n\in$                                                  & $\ket{0_L}$                 & $\ket{1_L}$   & Distance     & Measurement infidelity \\ \hline
\multirow{4}{*}{Fock} & \multirow{4}{*}{$\{0,1\}$} & \multirow{4}{*}{$\ket{0}$} & $\ket{2}$ & 2        & $(2.33\pm 0.03)\times 10^{-2}$                   \\ 
                                                                      &                    &                    &        $\ket{3}$           & 3                           & $(9.6\pm 0.6)\times 10^{-4}$                    \\ 
                                                                      &                    &                    &         $\ket{4}$           &                           4 & $(1.3\pm 0.2)\times 10^{-4}$                  \\ 
                                                                      &                    &                    &        $\ket{5}$            &                           5 & $\mathbf{(5.8\pm 1.3)\times 10^{-5}}$                   \\ \\ 
Binomial $(\hat{s})$                                                                & $\{1,2\}$                 &       $\ket{2}$           & $\frac{\ket{0}+\ket{4}}{\sqrt{2}}$                 & $2$                         & $(5.5\pm 0.2)\times 10^{-3}$                    \\
Binomial $(\hat{s},\hat{s}^\dag)$                                                                & $\{1,2,3\}$                 & $\ket{3}$ & $\frac{\ket{0}+\ket{6}}{\sqrt{2}}$                 & 3                 &  $\mathbf{(4.2\pm 0.2)\times 10^{-3}}$               \\ \hline
\end{tabular}
\caption{\label{tab:codes} \textbf{List of codes, mapping operations, and results.} Here we list the bosonic codes studied, along with their logical $Z$ codewords, the distance between codewords with respect to the lowering operator $a,$ and the subset of photon numbers for which the ancilla is excited. Experimental measurement infidelities are listed according to the logical encoding and quoted for the number $N$ of measurements which minimizes each. The best value for each class of code is highlighted in boldface.}
\end{table*}

In addition to the Fock-based codes described above, certain QEC codes can also be measured using the same procedure. The only difference in the implementation of the protocol is the choice of $S,$ which is summarized in Table~\ref{tab:codes}. In addition to the Fock-based codes, we also study two binomial codes \cite{article:binomial}. Such codes are based on superpositions of photon number states, and are designed to correct for different combinations of photon loss, photon gain, and dephasing errors. A universal gate set and error correction have been implemented for the lowest-order binomial code \cite{article:luyan_binomial}, and these codes have found application in quantum metrology \cite{article:heisenberg}. In this work, we do not consider codes in which any number state appears in both the $\ket{0_L}$ and $\ket{1_L}$ states. This means that the POVM of Eq.~\eqref{eq:povm} does not depend on phase for any of the $S$ studied. It is therefore sufficient to prepare number states as inputs to the measurement protocol and average the results appropriately.

Although majority voting is a convenient way to make assignments given a measurement record, it is not optimal.
In particular, the measurement fidelity worsens when too many readouts are incorporated into the majority vote, because of state relaxation. To make a uniform comparison between different codes, we use a Bayesian classifier (maximum likelihood estimator or MLE). Such a classification scheme is optimal and sufficiently general to classify any code \cite{article:hann}.
The results of such a classification scheme are shown in Fig.~\ref{fig:mle} as a function of the number of readouts taken. As more readouts are included in the classification, the infidelity improves monotonically, as expected for an MLE. Furthermore, the minimum infidelity improves as the distance of the encoding is increased. The final results are compiled in Table~\ref{tab:codes}, along with the definitions of $\ket{0/1_L}$ and $S.$ We note in particular that we have achieved a measurement fidelity of $1-6.5\times 10^{-5}\approx 0.9999$ when discriminating between states $\ket{0}$ and $\ket{5}$, and $1-4.2\times 10^{-3}\approx 0.996$ when discriminating between states $\ket{0_L}$ and $\ket{1_L}$ in the $S=2,N=1$ binomial code, both of which surpass the highest measurement fidelities reported in a circuit QED system \cite{article:walter}.

\section{Conclusions}
We have shown that multi-level encodings can be leveraged to improve measurement fidelity in quantum information systems. This idea was explored in two contexts. In the first set of experiments, the multiply excited states of a transmon qubit were used to protect measurements against errors due to state relaxation. Furthermore, it was shown how this technique can be used to mitigate SPAM errors. In the second set of experiments, a repeated readout scheme was used to measure qubits encoded in a harmonic storage mode. In this scheme, repeated readouts suppress the effect of mapping and readout errors, and distance between codewords suppresses the effect of state relaxation. Using an adaptive reset scheme relying on realtime feedforward logic, this scheme was implemented with a single ancilla.

Measurement is a crucial resource in quantum information processing. In particular, reliable measurements are necessary for high-fidelity teleported operations \cite{article:chou}, which are an important component of a modular architecture for quantum computation \cite{article:modular}. Therefore, we expect our results to find applications in a variety of future experiments.

\begin{acknowledgments}
The authors wish to thank C.~J. Axline for assistance with sample preparation during an early stage of this experiment, N. Frattini and K.~Sliwa for providing the
Josephson parametric converter, and N.~Ofek for providing the logic for the FPGA control system used in this experiment. 
Facilities use was supported by the Yale SEAS cleanroom and YINQE.
This research was supported by the U.S. Army Research Office grants W911NF-18-1-0212 and W911NF-16-1-0349, and by the AFOSR MURI (FA9550-14-1-0052). L.J. acknowledges support from the Packard Foundation (2013-39273).
C.T.H. acknowledges support from the NSF GRFP (DGE1752134).
\end{acknowledgments}

\clearpage
\bibliography{reference_list}

\end{document}



\title{Supplemental material for ``High-fidelity measurement of qubits encoded in multilevel superconducting circuits''}





\pacs{}

\maketitle



\section*{Readout QND-ness}
\begin{figure}[h]
\includegraphics{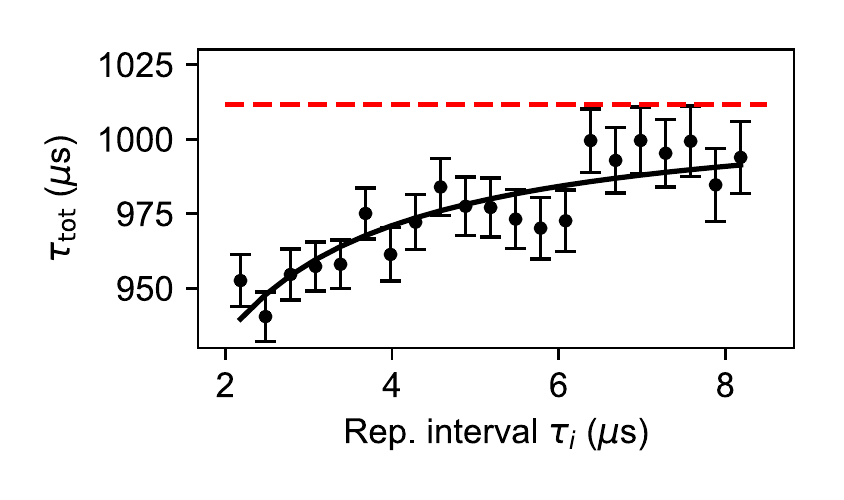}
\caption{\label{fig:qndness} \textbf{QND-ness of ancilla readout.} Cavity $T_1$ measurements were performed with ancilla readouts during the delay repeated with a variable interval time $\tau_\mathrm{i}.$ The data are fit to a model (solid line) $1/\tau_\mathrm{tot}=1/\tau_0+P_D/\tau_\mathrm{i}.$ From this we infer a demolition probability per readout of $P_D=0.02\%,$ corresponding to a QND-ness of $99.98\%.$ The fit parameter $\tau_0=1.01$~ms is indicated by a dashed red line.}
\end{figure}

Based on the low-energy Hamiltonian of a transmon coupled to harmonic oscillators~\cite{article:bbq}, we do not expect the measurement protocol to induce additional relaxation in the storage mode. However, several experimental and theoretical works have explored readout-induced state transitions in a transmon~\cite{article:siddiqi_T1nbar,aps:T1vsnbar,article:blais_T1nbar,article:sank}. In the context of harmonic storage oscillators, one study has shown that parity measurements, while highly quantum non-demolition (QND), can induce a small amount of additional relaxation \cite{article:parity}. In that experiment, the total relaxation rate was modeled as a combination of the bare storage lifetime $\tau_0$ and a demolition probability $P_D$ associated with each parity measurement. Varying the rate, $1/\tau_\mathrm{i},$ at which repeated parity measurements were performed, allowed $P_D$ to be measured. Here, we perform a similar calibration to extract the additional rate of photon loss from the storage induced by a transmon readout. As shown in Fig.~\ref{fig:qndness}, we find that transmon readout is $99.98\%$ QND with respect to the storage mode.

\section*{Maximum likelihood estimator}
The repeated readout protocol is modeled as a hidden Markov model, in which the underlying states represent the storage mode photon number, transitions between hidden states represent photon loss and gain, and emission matrices represent error-prone readouts with the ancilla. If the transition and emission matrices $T$ and $E$ are known, then a given series of emissions (readouts) can be converted into a best estimate of the initial state of the storage mode~\cite{article:hann}.

To determine the measurement fidelities shown in
\ifarxiv Fig.~6\else Fig.~\ref{fig:mle}\fi, the forward-backward algorithm \cite{book:nr} is applied on a single-shot basis to every sequence of readout outcomes. The transition matrix used at each step takes into account the total time taken by that cycle's reset operation. For each readout sequence, a degree of belief $p'(n_0)$ for the initial number of photons in the storage is calculated. The prior distribution $p(n_0)$ is based on the code being measured. These probabilities are then converted into a logical measurement outcome in the natural way.
For example, when distinguishing $(\ket{0}+\ket{4})/\sqrt{2}$ and $\ket{2},$ the prior distribution is $p(0)=p(4)=0.25, p(2)=0.5$; the assignment is ``$0_L$'' if $p'(0)+p'(4)>p'(2)$ and ``$1_L$'' otherwise.

\section*{State preparation}
\label{sec:sup_prep}
As shown in \ifarxiv Fig.~4\else Fig.~\ref{fig:schematics}\fi, we use a heralded method of state preparation in order to characterize the repeated measurement protocol. Here we investigate whether this method will affect the experimentally determined measurement fidelities. Specifically, we show that (i) the heralded states have high fidelity, and (ii) the main source of error is photon loss during the final check measurement. This means that although the state preparation is imperfect, we will only rarely prepare states in $S$ which are meant to be in $\bar{S},$ or vice-versa. Furthermore, the error is of the form predicted by the theoretical description of the protocol \cite{article:hann}. Therefore, we do not need to explicitly account for the preparation error in analyzing our experiments.

\begin{figure}[h]
\includegraphics{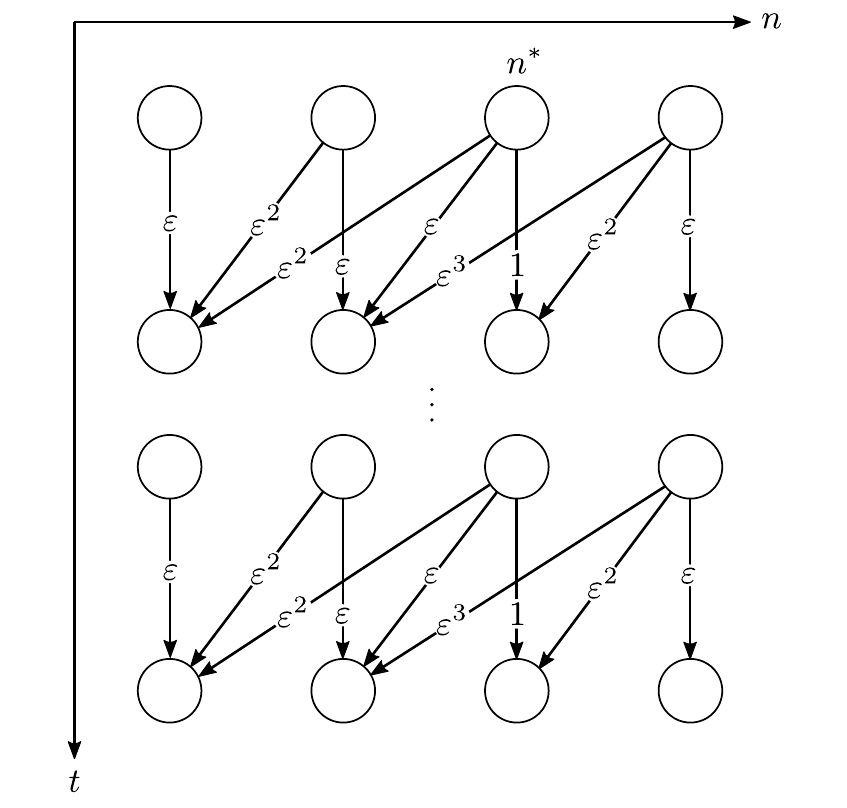}
\caption{\label{fig:preparation} \textbf{Illustration of preparation errors.} The probabilities $P_t(n)$ are represented by the rows and columns of the network above. We imagine each node to contain the respective probability. $P_t(n)$ can be calculated by applying a linear map to the previous probabilities, $P_{t-1}(n),$ at each time step. These linear operations are shown as edges in the graph. By tracing out paths through the network, we can visualize the important terms in the expansion \eqref{eq:prepared_dob}.}
\end{figure}

Our approach is to consider the degree of belief for the current state of the storage mode as additional checks are passed. We suppose that at each step of the verification, there is first a check measurement, and then the possibility of a state transition. Denote the probability of passing a check measurement when in state $\ket{n}$ by $E_n,$ and the probability of transitioning from state $\ket{n}$ to state $\ket{n'}$ as $T_{n\rightarrow n'}.$ Then by keeping track of $P_t(n)\equiv P(\textrm{currently }n~\cap~ t \textrm{ checks passed})$ as measurements are added, we can update it at every step:
\begin{equation}\label{eq:prepared_dob}P_{t+1}(n')=\sum_n P_t(n)E_n T_{n\to n'}.\end{equation}
This function is proportional to the conditioned degree of belief, $P(\textrm{currently }n~\mid~t\textrm{ checks passed}).$ Our claim is that once several checks have passed, this state is nearly the ideal photon number state being prepared, but subject to state relaxation during the last check measurement. To see why, consider the tree shown in Fig.~\ref{fig:preparation}, whose branches at every step represent the terms in the sum above. $\varepsilon$ denotes a generic error probability (either a false positive of the check measurement, or the time per check divided by the storage lifetime). We see that to first order, only two paths contribute to $P_t(n).$ First, there is the ideal path, in which the desired number state  $\ket{n^*}$ was prepared correctly and confirmed. Second, there is the possibility that the correct state was prepared and confirmed, but that it decayed to $\ket{n^*-1}$ at the very end of the protocol.

The argument above is robust to the details of the number state creation and check measurements. Also, we stress that the remaining preparation error, photon loss, is the very error which our measurement protocol is robust to.

\section*{Reset errors}
\label{sec:sup_reset}
\begin{figure*}
\includegraphics[width=\textwidth]{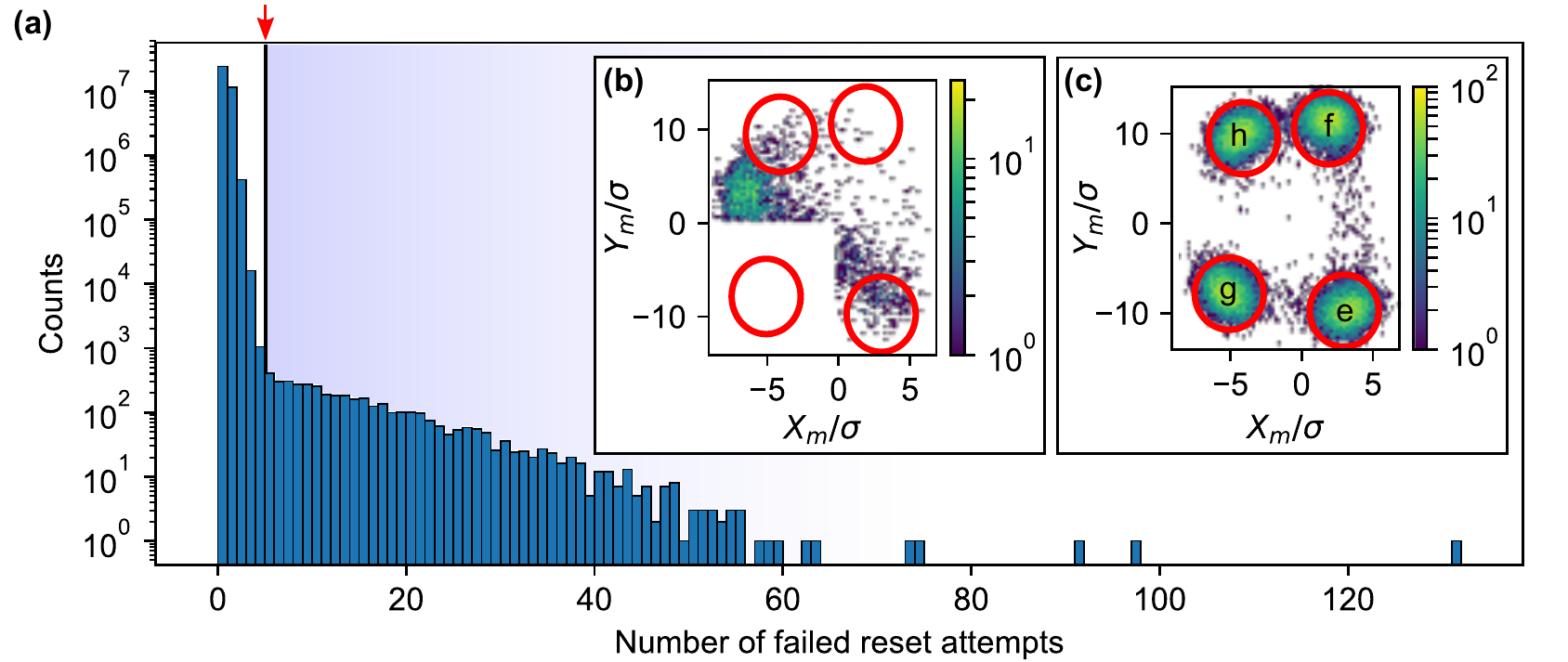}
\caption{\label{fig:stuck} \textbf{Ancilla reset failures.} \textbf{(a)} Histogram of the number of feedforward operations which are required to reset the ancilla. The red arrow near the top-left indicates the threshold we use to declare the ancilla ``stuck.'' Of 1.8 million repeated measurement experiments, only 3,058 instances (or $0.2\%$) had a stuck ancilla during any reset operation. \textbf{(b)} Histogram of readout signals corresponding to a stuck ancilla. Because reading out the ancilla in its ground state is the success condition for the reset operation, there are no counts in the ``g'' quadrant. \textbf{(c)} In a separate experiment, states $\ket{g}$ through $\ket{h}$ are prepared in the ancilla and measured. The corresponding readout signals are shown for reference. The red circles are a guide for the eye.}
\end{figure*}

The predictions of \ifarxiv Eq.~(4) \else Eq.~\eqref{eq:theory} \fi assume that the ancilla is perfectly reset every repeated measurement cycle, and that the reset operation takes a fixed amount of time. While these are good approximations, they break down when we are interested in very small sources of error. Fig.~\ref{fig:stuck}(a) shows a histogram of the number of attempts which were required in order for the ancilla reset operation to complete. As expected, the reset almost always succeeds in placing the ancilla in its ground state within a few conditional operations. However, there is an unexpected tail in the distribution: this tail represents rare instances in which tens of consecutive readouts and conditional pulses failed to put the ancilla in its ground state. We refer to the ancilla as being ``stuck'' if the reset operation takes five or more iterations to succeed. These ``stuck'' instances can be understood as the result of population in the fifth or higher transmon level, which would not be handled properly by the conditional logic used. The detailed mechanism of these excitations is not apparent from our measurements. Fig.~\ref{fig:stuck}(b) shows a histogram of all measurements $m$ after which the reset was ``stuck'' (as defined in \ifarxiv Fig.~Fig.~4(a)\else \ref{fig:schematics}(a)\fi), revealing the presence of population in higher states. As a reference, a histogram of integrated measurement signals corresponding to the first four levels of the transmon is shown in Fig.~\ref{fig:stuck}(c), with red circles as a guide to the eye.

\section*{System parameters}
\begin{table*}
\begin{tabular}{ccc}\hline\hline
Parameter & Convention & Value \\ \hline
Storage frequency & $\omega_s\hat{s}^\dagger \hat{s}$ & $\omega_s/2\pi=4.5$~GHz \\
Ancilla frequency & $\omega_s\hat{t}^\dagger \hat{t}$ & $\omega_t/2\pi=4.2$~GHz \\
Readout frequency & $\omega_r\hat{r}^\dagger \hat{r}$ & $\omega_r/2\pi=9.33$~GHz\\
\\
Ancilla anharmonicity & $\frac{\alpha}{2}\hat{t}^\dag\hat{t}^\dag \hat{t}\hat{t}$ & $\alpha/2\pi=-137$~MHz \\
Storage anharmonicity & $\frac{K_s}{2}\hat{s}^\dag\hat{s}^\dag\hat{s}\hat{s}$ & $K_s/2\pi=-2.2$~kHz \\
Storage-ancilla interaction strength & $\chi_{st}\hat{s}^\dag\hat{s}\ket{e}\bra{e}$ & $\chi_{st}/2\pi=-900$~kHz \\
\\
Storage lifetime & $\frac{1}{T_1^{(s)}}\mathcal{D}[\hat{s}]$ & $T_1^{(s)}=1/\kappa_\downarrow=0.99$~ms \\
Storage thermal population & & $\bar{n}_s\approx 2.1\%$\\
Ancilla $e$ lifetime & $\frac{1}{T_1^{(ge)}}\mathcal{D}[\ket{g}\bra{e}]$ & $T_1^{(ge)}\approx 51~\mu$s \\
Ancilla $f$ lifetime & $\frac{1}{T_1^{(ef)}}\mathcal{D}[\ket{e}\bra{f}]$ & $T_1^{(ef)}\approx 47~\mu$s \\
Ancilla $h$ lifetime & $\frac{1}{T_1^{(fh)}}\mathcal{D}[\ket{f}\bra{h}]$ & $T_1^{(fh)}\approx 40~\mu$s \\
$\ket{g}+\ket{e}$ coherence time &  & $T_2^{(ge)}\approx 74~\mu$s\\
$\ket{g}+\ket{f}$ coherence time &  & $T_2^{(gf)}\approx 57~\mu$s\\
Ancilla thermal population & & $\bar{n}_t\approx 0.4\%$\\ \hline

\end{tabular}
\caption{\label{tab:parameters}\textbf{Measured parameters of the cQED system.} Hamiltonian terms are given divided by $\hbar.$ Our convention for the Lindblad superoperator is $\mathcal{D}[\hat{L}]\rho=\hat{L}\rho\hat{L}^\dagger-\frac{1}{2}\{\hat{L}^\dagger \hat{L},\rho\}.$}
\end{table*}
Measured system parameters are summarized in Table~\ref{tab:parameters}. The parameters used in the theory curves in \ifarxiv Fig.~5 \else Fig.~\ref{fig:votes} \fi of the main text were determined as follows: $\delta_0=5.2\times 10^{-2}$ and $\delta_1=1.5\times 10^{-3}$ were taken as the first point in the $0$-photon and $5$-photon curve, respectively. $\kappa_\uparrow\tau=2.7\times 10^{-4}$ was obtained by fitting to the $1$-photon curve. $\kappa_\downarrow\tau$ should be regarded as a single effective parameter calculated as follows. The duration of the mapping pulse is is $t_\mathrm{map}=2.4~\mu$s, and one ancilla readout and reset attempt have a duration of $t_\mathrm{r}\approx 2.16~\mu$s. $T_1^{(s)}$ was determined by independent calibration and combined with the measured QND-ness to obtain $\kappa_\downarrow\tau=(t_\mathrm{map}+t_\mathrm{r})/T_1^{(s)}+P_D=4.8\times 10^{-3}.$

\section*{QuTiP simulation details}
In \ifarxiv Sec.~III \else Sec.~\ref{sec:transmon} \fi of the main text, we describe simulations of $g$--$e$ and $e$--$f$ $\pi$-pulses. These were performed using the anharmonicity, relaxation times, and coherence times in Table~\ref{tab:parameters}.
The jump operators simulated were
\begin{align}
\frac{1}{T_1^{(ge)}}\mathcal{D}[\ket{g}\bra{e}]\rho+\frac{1}{T_1^{(ef)}}\mathcal{D}[\ket{e}\bra{f}]\rho\\
+2\gamma_\phi^{(e)}\mathcal{D}[\ket{e}\bra{e}]\rho+2\gamma_\phi^{(f)}\mathcal{D}[\ket{f}\bra{f}]\rho,\nonumber
\end{align}
where $\gamma_\phi^{(e)}=1/T_2^{(ge)}-1/2T_1^{(ge)}$ and $\gamma_\phi^{(f)}=1/T_2^{(gf)}-1/2T_1^{(ef)}.$
The parameters of the pulses are listed in Table~\ref{tab:pulse_parameters}. Several amplitudes were simulated for each pulse, and the one maximizing the $g\to e$ or $e\to f$ probability was chosen in each case. In this way we can calculate $P_g$ and $P_g^\textrm{shelved}$ quoted in the main text. From the simulated populations in states $\ket{g},$ $\ket{e},$ and $\ket{f}$ immediately after the shelving pulse, $P_g^\textrm{meas}$ was calculated by solving the differential equations
\begin{align}
\dot{p}(t)=\begin{bmatrix}0 & +1/T_1^{(ge)} & 0 \\ 0 & -1/T_1^{(ge)} & +1/T_1^{(ef)} \\ 0 & 0 & -1/T_1^{(ef)}\end{bmatrix}p(t)
\end{align}
for the vector $p(t)$ of populations.
\begin{table}
\begin{tabular}{cc}\hline\hline
Parameter & Value \\ \hline
Envelope width & $\sigma_{ge}=5$~ns \\
 & $\sigma_{ef}=6$~ns\\
 \\
Pulse length$/\sigma$ & $N_{ge}=8$\\
& $N_{ef}=6$\\
\\
Drive detuning & $\Delta_{ge}=3.899$~MHz \\
& $\Delta_{ef}=2.67$~MHz \\ \hline

\end{tabular}
\caption{\label{tab:pulse_parameters}\textbf{Pulse parameters used in shelving experiment.} The pulses were chosen with Gaussian envelopes and experimentally determined detunings. The drives are blue-detuned with respect to the respective transitions.}
\end{table}

\section*{Simultaneous number-selective pulses}
The pulse which maps information from the storage mode onto the ancilla can be thought of as an $S$--controlled $\pi$-pulse. That is, we want to enact the following unitary:
\begin{align}
\hat{U}_\mathrm{map}&=\sum_{\ket{n}\in S}\ket{n}\bra{n}\Big(\ket{g}\bra{e}+\ket{e}\bra{g}+\ket{f}\bra{f}+\cdots\Big)\\
&+ \sum_{\ket{n}\not\in S}\ket{n}\bra{n}\Big(\ket{g}\bra{g}+\ket{e}\bra{e}+\ket{f}\bra{f}+\cdots\Big).\nonumber
\end{align}
This gate was implemented using simulataneous gaussian pulses centered around the appropriate qubit frequencies and truncated to $\pm 2\sigma_t$. The envelopes were chosen with a standard deviation of $\sigma_t=600~\mathrm{ns},$ corresponding to a frequency width of $\sigma_f=1/2\pi\sigma_t=265~\mathrm{kHz}.$ This choice balances the pulse length (which should not be too long, to avoid storage mode relaxation and ancilla decoherence) against the pulse bandwidth (which must be small compared with the dispersive shift in order to be number-selective). The pulses are applied with empirically chosen detunings in order to achieve a reasonably high mapping pulse fidelity. The mapping is calibrated separately for each choice of $S$ (defined in \ifarxiv Table~I \else Table~\ref{tab:codes} \fi of the main text).

\section*{Experimental setup}
The cQED package used in the experiment is mounted to the base stage of a dilution refrigerator. Drives are single-sideband modulated using an FPGA system and delivered to the device after appropriate filtering and attenuation. The outgoing readout signal is amplified by a JPC at base temperature, a HEMT at 4~K, and a MITEQ amplifier at room temperature. It is then mixed down to 50~MHz and recorded by an ADC. The experimental details are essentially the same as those described in Ref.~\cite{article:ft}.

\bibliography{reference_list}